# Effective Voice: Beyond Exit and Affect in Online Communities

Seth Frey[1], Nathan Schneider[2]

[1] Communication, University California Davis. sethfrey@ucdavis.edu

[2] Media Studies, University Colorado Boulder. nathan.schneider@colorado.edu

**Abstract**

*Online communities provide ample opportunities for user self-expression but generally lack the means for average users to exercise direct control over community policies. This paper sets out to identify a set of strategies and techniques through which the voices of participants might be better heard through defined mechanisms for institutional governance. Drawing on Albert O. Hirschman's distinction between "exit" and "voice" in institutional life, it introduces a further distinction between two kinds of participation: effective voice, as opposed to the far more widespread practices of affective voice. Effective voice is a form of individual or collective speech that brings about a binding effect according to transparent processes. Platform developers and researchers might explore this neglected form of voice by introducing mechanisms for authority and accountability, collective action, and community evolution.*

**Effective Voice: Beyond Exit and Affect in Online Communities**

"A riot is the language of the unheard," Martin Luther King, Jr., famously said on several occasions of uprisings in US cities. The prophet of nonviolence's sympathetic interpretation of smashed windows and burning neighborhoods was an indictment of a United States that understood itself as a liberal democracy—the kind of society meant to excel at hearing its citizens, at having a proper channel for every grievance and desire. It should similarly come as an affront to the Internet pioneers' widely held norms of free speech and permissionless participation that network denizens do not perceive their voices to be heard, and are acting out accordingly. Whitney Phillips (2015, 12), for instance, came to recognize her study of online trolling to be "as much a critique of dominant institutions as it is of the trolls who operate within them"—for deviance mirrors back what institutions ignore or attempt to suppress. A study of the "AMAgeddon" revolt among Reddit users observes, "Users made it clear that they were no longer content with supplying and moderating all of Reddit's content; they also wanted a voice in the governance and policy determinations impacting the community" (Centivany and Glushko 2016). By these accounts, both urban rioting and online miscreantry have at their source the failure of institutions to hear. As online communities become ever more implicated in patterns of misinformation, abusive behavior, and declining trust in public institutions, their failures to self-regulate have ever larger consequences.

This paper sets out to identify a set of strategies and techniques through which the voices of participants in online communities might be better heard through defined institutional mechanisms. A growing chorus of scholars has pointed to the absence of and the need for such mechanisms. Seering et al. (2019) found, in a study of dozens of moderators across three social media platforms, that decisions "are often made without substantial feedback from non-moderators"; the authors propose that "platforms might consider developing features that allow all users to get involved in self-governance if they so choose." A study that mapped "dimensions of contemporary participation" in online contexts detected, among the major platforms, "an inability or unwillingness to incorporate users' input as to what the activity should be or how it should proceed" (C. Kelty et al. 2015). Its authors further speculated on the consequences for civic life in general: "Would it not be safe to assume that these forms of participation are patterning people in ways that dispose them to some kinds of participation but not others?"

We introduce a distinction between two kinds of participation in self-governance: *effective voice*, as opposed to *affective voice*. True to the contours of affect theory generally (Ahmed 2004; Massumi 1995; Papacharissi 2015), affective voice is expression that courses through a collective cultural and emotional landscape, seeking to move, motivate, and mobilize its hearers. This is the type of speech that commercial social-media platforms primarily enable. Effective voice is rather straightforward in contrast, but comparatively rare on Internet platforms: *a form of individual or collective speech that brings about a binding effect according to transparent processes*. If affective voice is the debate, effective voice is the vote; if affective voice expresses and persuades, effective voice determines. In what follows, we set out to probe the absence and possibilities of effective voice in the peer governance of online communities. Although our distinction is akin to Sherry Arnstein's "ladder of citizen participation" (Arnstein 1969;

Carpentier 2016; C. M. Kelty 2019), with its hierarchy of effective "citizen control" over affective "tokenism," we do not regard one as *a priori* superior to the other. A case for effective voice may constitute an implicit critique of affective expression, deliberative democracy, and other popular celebrations of traditional voice, but we ultimately view effect and affect as complementary.

Many readers will recognize the idiom of *voice* as stemming from economist Albert O. Hirschman's (1970) famous distinction between exit and voice—the possible recourses, as he understood them, for people dissatisfied with their interactions with companies, organizations, and states. We count among the legions of scholars across many fields who have found Hirschman's simple dyad to be generative. In mapping out the promise of effective voice, we draw on Hirschman and a selection of his intellectual descendants, who have added to and sharpened his capacious schema. We also put this legacy into conversation with another tradition of organizational theory at the edges of economics: the Institutional Analysis and Design school, stemming from the work of Elinor Ostrom and her circle.

We begin by reviewing the insights of Hirschman and others before us who have drawn on him, particularly identifying the varieties of participant agency and their assorted usefulness. Next, we turn these typologies on the Internet age, assessing the relevance of Hirschman and his successors for the governance of online communities—highlighting in particular these communities' enduring reliance on exit and the corresponding paucity of effective voice. With that analysis in hand, we then identify possible mechanisms for effective voice in social platforms. We by no means offer these technological solutions to replace the interpersonal

consensus- and culture- building work typical of strong online communities, rather we introduce them to complement, support, or better engage community members in this kind of collective meaning making. These mechanisms can present opportunities such as future research experiments, user-experience designs for platform developers, and demands that organized users might make of their platforms.

Our focus is on users of online communities—peers who inhabit roughly comparable status and who self-manage aspects of their shared virtual spaces. We consider more exalted layers of Internet governance, such as the platform companies or network protocols, mainly to the extent that they organize and orchestrate communities among users. But we principally direct our attention to the opportunities for voice that ordinary users might experience in their daily online lives.

## 1. Exit, voice, and agency

By the time he wrote *Exit, Voice, and Loyalty*, Hirschman had absorbed much experience with both departures and struggles for change from within. He had fled several countries (including his native Germany) and fought in several armies (against Nazi Germany and its allies). He floated from post to post in the US government and international agencies, and distinguished himself as a heterodox development economist. It may be that his repeated experience of compelled exits motivated a longing for voice and for the possibility of loyalty.

Hirschman first presents the exit-voice distinction in the context of companies producing consumer goods, but before long it becomes clear that he sees the distinction's relevance much more broadly in social and political life. The business-like orientation was a gambit, as he later put it, "to convince economists of the importance and usefulness, for the analysis of economic phenomena, of an essentially political concept such as voice" (Hirschman 1980, 431). Exit, for him, is when "some customers stop buying the firm's products or some members leave the organization"; voice is when "the firm's customers or the organization's members express their dissatisfaction directly to management or to some other authority" (Hirschman 1970, 4).

The options of exit and voice are not mutually exclusive. Both can be "overdone" (31), and throughout the book he probes their curious interplay—how voice might be weakened by some sorts of exit, for instance, but strengthened by others—leaving those in search of an easy formula with only a warning of "the inborn tendencies toward instability of any optimal mix" (126).

The context of loyalty enters later as a mediating factor for understanding "the conditions favoring the coexistence of exit and voice," one which "holds exit at bay and activates voice" (78). Participants' access to voice—particularly voice that they perceive as meaningful—fosters loyalty, which diminishes the likelihood of exit. Even when exit is an available option, that is, "a member who wields (or thinks he [sic] wields) considerable power in any organization … is likely to develop a strong affection for the organization." Loyalty also helps produce "slack," which Hirschman described as an essential reserve of institutional capacity that can be mobilized during times of difficulty or expansion (14). Hirschman's concept of loyalty represents an enticement for organizations to overcome their disinclination toward voice (92-93) and see cause for embracing it.

Hirschman's conception of voice is largely in the terrain of the affective, particularly concerning consumer-corporate relations, where no effective mechanisms are typically present. Yet he treats both consumer complaints and involvement in electoral politics as forms of voice, suggesting that rhetorical pleas and voting mechanisms share common, potentially loyalty-inducing qualities. We appreciate the value of recognizing the affective and effective together as forms of voice, yet we will contend that their differences matter as well.

**Exit and voice since Hirschman**

*Exit, Voice, and Loyalty*'s reception within economics was cool—a review by economist Gordon Tullock (1970) characterized its central concern as "not monumental"—but the concepts of exit and voice were quickly picked up elsewhere. Since its introduction, the exit-voice distinction has found place in media studies (Flew 2009), urban studies (Young 1974), labor studies (Saunders 1992), migration studies (Newland 2010), the marketing and consumer literatures (Raval 2019), group behavior (Tajfel 1975), deliberative democracy (Gastil 2000), and the social psychology of romantic relationships (Rusbult, Zembrodt, and Gunn 1982). There is also a considerable exit-voice literature surrounding Internet cultures (Centivany and Glushko 2016; C. Kelty et al. 2015; Currie, Kelty, and Murillo 2013; Lu et al. 2013; Marlin-Bennett and Thornton 2012; Martin, Parry, and Flowers 2015; Nunziato 2000; Rezabakhsh et al. 2006; Kucuk 2008; Michaelsen 2018), which we will turn to more directly in the next section.

Throughout this *ouvre*, Hirschman's distinction has proven relevant across diverse fields of institutional design. Scholars have mapped exit and voice to distinctions between market and state (Somin 2011), neoliberal and democratic ideologies (Lowery, DeHoog, and Lyons 1992;

Hirschman 1980), or federalist and unitary government (Young 1976), for example. Loyalty's role has been just as varied (J. W. Graham and Keeley 1992), appearing both as a third concept on equal footing with exit and voice or, as here, in a supporting position that elaborates the relationship between the other two. And whereas Hirschman stressed the value of voice, most interpretations have run with the neutrality of his book's title, which seems to give exit equal footing in a balanced comparative analysis of the varieties of agency.

Elaborations of exit and voice consistently distinguish between the rulers and the ruled, the insiders and outsiders, the actors and reactors. The disgruntled customer and the frustrated citizen alike stand at the mercy of remote decision-makers. Starting from Hirschman's original frame of dissatisfaction, analyses of exit and voice have stressed the agency of grievance, as opposed to that of more positive, constructive forms of action. By and large, however, the rigid insider-outsider distinction is appropriate to the domains at hand, such as in a conventional corporation's relationships with customers and employees or a state's relationships with citizens. In many of these domains, also, the forms of voice available are chiefly affective—resulting in redress only when the decision-makers opt to listen. Affective mechanisms may suffice for most people outside of authority positions: late in life, Hirschman celebrated merely "having opinions" for its ability to contribute to human "well-being" (Hirschman 1989). Yet in some important cases the lines of authority are porous; outsiders do—and should—have direct roles in decision-making.

In election campaigns, voluntary associations, religious groups, and worker cooperatives, for example, participants can access modalities of voice that function as direct levers of power.

These might include ballots, petitions, board seats, and other positions of authority. And while individual influence may be small and fractional, collective action on these levers can be sufficient to compel action at the highest levels of the organizations in question. In contrast to affective expressions that seek to persuade power holders, these more direct interventions are what we call effective voice.

Consider, also, Heather K. Gerken's analysis of the varieties of agency available to low- and mid-level bureaucrats, the front-line agents of states and other large organizations (Gerken 2013). Bureaucrats may stand just as far from the positions of power as any other citizens, but they do hold power, the power to be selective in their interpretation and implementation of policy directives from above. She describes their foot-dragging or outright refusal as "disloyalty," and yet it is also a feature of the governance process that enables citizen input and control. Such acts could thus fall under the purview of effective voice.

To be clear, the concepts of exit and voice are not airtight classifications. The same goes for our distinction between effective and affective voice. Building on Hirschman's own reflections, later studies have contested the lines between exit and voice (Taylor 2016; Somin 2011), presented examples of their complex interplay (Gofen 2012; Michaelsen 2018), and highlighted forms of agency that neither is adequate to describe (Farrell 1983; Gerken 2013). The ambiguity of the framework probably goes hand-in-hand with its broad applicability, as well as its capacity to convey insights across fields of study (Ellerman 2005).

## 2. Exit, voice, and online communities

What would Hirschman have said differently if he had been writing *Exit, Voice, and Loyalty*—as we write this—in Google Docs? Hirschman worked on the book while a fellow at the Center for Advanced Study in the Behavioral Sciences, perched on a hill overlooking Stanford University and what would come to be called Silicon Valley, home to many of the world's largest online social platforms. Yet understandably his analysis did not anticipate how platforms of the future would alter the interactions among his core concepts.

In comparison to the industrial firms that Hirschman considered, devising and selling a product according to a "pipeline" model, commercial online platforms act as points of connection among users, who may or may not resemble customers (Parker, Alstyne, and Choudary 2016). A person might use a social-media platform for many years without paying anything for the service, while "brands" pay the platform company to access that user's attention. There is a layer of abstraction that separates the users' interactions on the platform from the economics of the platform. For example, when a user departs in disgust from one community on a given platform, this apparent exit may provide net benefit for the platform company, which can then incorporate that departure into a more accurate profile of the user for advertisers. Platforms thus tame the business consequences of user exit. Hirschman (1970, 40) also assumed that firms are plentiful and civil society organizations are rare, while the opposite seems to be the case in online life. Interest groups proliferate easily, and one is not constrained to choose among only those that have established local chapters and storefronts; meanwhile, network effects ensure that relatively few large Internet companies tend to dominate any given vertical (Belleflamme and Peitz 2018). The result, paradoxically, is both more choice and more consolidation—lowering the floor to participation but raising the ceiling of control. Catherine J. Turco (2016) has shown how this pattern has extended back into the management of platform companies, where open, affective "conversation" proliferates among employees but only reinforces the hierarchies of effective decision-making.

Here we highlight two general tendencies of the online context that, in comparison to the cases with which Hirschman was familiar, renegotiate the interactions between exit and voice.

**Lower agency costs**

Online social platforms make certain forms of both exit and voice easier, together with other forms of agency. In Hirschman's time, consumers exercising voice against a company might require a letter-writing campaign, a boycott backed by national organizations, or a Ralph Nader-style charismatic activist; exiting a local civic organization could require time-consuming paperwork. Online, ostensibly, being heard requires only a viral tweet that mentions the company's handle; leaving a group requires only pressing the "Leave Group" button.

Scholars have long observed the ease of exit from online communities as a general pattern (Currie, Kelty, and Murillo 2013; D. R. Johnson and Post 2001; Nunziato 2000). The more prosaic exit can take several forms. One form is what Gofen (2012) describes as "entrepreneurial exit"—when users "exit proactively by creating a viable alternative themselves." Creating a new email list, forum, or other community domain frequently requires just a few steps and no monetary cost on popular platforms (Kraut, Resnick, and Kiesler 2011). For users with the relevant skills, entire programs with free/open-source code can be "forked" with a single terminal command, producing an exact copy for modification and redeployment (Nyman and Lindman 2013). Although convincing other users to join and use new communities or software instances may prove difficult, the low-to-nonexistent costs of creation assist in making "a viable alternative" online easier than, for instance, incorporating a legal entity or writing bespoke software from scratch. Participating in and maintaining communities may also present lower barriers than offline affinity organizations; interactions can occur through asynchronous interaction rather than requiring in-person meetings, and a community's space can persist on some commercial platforms with little or no user maintenance. Several studies, drawing on Hirschman, have celebrated the newfound consumer exit-power with respect to firms, thanks to the proliferation of online price and quality information (Kucuk 2008; Rezabakhsh et al. 2006; Gans, Goldfarb, and Lederman 2017; Kumar, Qiu, and Kumar 2018).

Online platforms seem to facilitate voice just as well as exit. They offer the potential, at least, for any user to wield outsized influence through "viral" messages that spread across a network. Users' messages can achieve global reach that would have been unavailable before the Internet. Low-cost voice has produced changes in behavior in realms such as consumer products (Kucuk 2008; Rezabakhsh et al. 2006), government services (Meijer 2007), employee grievances (Martin, Parry, and Flowers 2015), and protest movements (Tufekci 2017), to name a few. The above-mentioned revolt among Reddit users saw moderators engage in concerted action that succeeded in deposing a CEO (Matias 2016; Centivany and Glushko 2016). Platforms have enabled the rise of powerful and persistent counterpublics among marginalized people, such as "Black Twitter" (R. Graham and Smith 2016). Practices of hacking and trolling exploit features of the digital environment to spread messages in novel ways (Coleman 2012). Perhaps it is due to the dizzying array of opportunities for voice online that users might fail to notice what opportunities are not on offer, or the nearly ubiquitous regimes of control that surround them.

**Consolidated control**

Hirschman (1970, 60) raised the specter of the "lazy monopolist," "who thrives on the limited exit possibilities existing for its most fastidious and well-to-do customers." While the topology of online communities offers ample exit options of certain kinds, it also produces new sorts of monopolists.

Undergirding the low-cost forms of agency that networked cultures enable are unprecedented regimes of control. At the base layer lies the protocol, the technical rules that all network participants must follow in order to participate (Galloway 2006). Atop the protocols stand platforms—the consumer-facing sites and apps—whose collective name evokes a flat neutrality belied by their unprecedented power and commercial agendas (Gillespie 2010). The communities that reside on platforms, atop protocols, exhibit similar patterns of control as the systems that undergird them. Just as the competition among protocols and platforms tends to be "winner take all," user communities tend to cluster according to a "fat tailed" distribution: a few very large groups looming over a large number of very small ones (Shirky 2003), while within communities a small number of members tend to wield outsized influence (S. L. Johnson, Faraj, and Kudaravalli 2014). The tendency for centralized outcomes despite open networks and cheap agency have also been theorized in terms of "network effects" (Belleflamme and Peitz 2018) and experienced by users as "lock in" (J. P. Johnson 2020); early pioneers of open, decentralized Internet technologies typically did not expect, and now mourn, these outcomes (Schneider 2019).

Atop the consolidated regimes of platforms and the communities within them, those in positions of power can carry out decisive and granular acts of control. Stemming from computer access-control systems and norms that arose on early online networks, a logic of "implicit feudalism" grants administrators the power to silence and remove users, often unilaterally (Schneider 2020). Platform designers apparently do not presently deem more participatory mechanisms of control to be necessary to retain or attract users. Prevailing modes of lock-in may be dangerously sufficient. The historical record suggests that when societies lack meaningful exit options, there has been little pressure for those who hold power to share it (Stasavage 2020). Administrators of various online communities nevertheless expend extensive affective labor—generally uncompensated—to foster perceptions of accountability (Huffaker 2010; Seering et al. 2019), while holding effectively dictatorial powers over their peers. Yet the presence of consolidated control may be less perceptible than the availability of cheap exit and voice, giving rise to a dominant discourse of freedom and limitless possibility (Rheingold 1993).

**Limits of exit and affect**

Hirschman (1970, 83) warns that, in a healthy organizational design, "there should be the possibility of exit, but exit should not be too easy or too attractive as soon as deterioration of one's own organization sets in" (83). Online communities have a confounding relationship to this advice. On the one hand, exit from a community is typically a click away, making exit ostensibly simple. On the other hand, exiting platform infrastructures can be painfully difficult, as sociologist Janet Vertesi found when she sought to prevent large tech companies from knowing about her pregnancy; the task involved disrupting relationships, utterly reorganizing her digital life, and ultimately constructing her own mobile phone (Matias 2018). With regard to peer-production communities, as Nadia Eghbal puts it, "Forking is a technical right. Socially, it's much harder to execute" (Stacoviak and Santo 2020). Exit is not a real option for most community members. Social pressures might similarly make even a Facebook Group's easy exit buttons less (affectively) easy to press than they (effectively) appear. When there is not "the possibility of exit," meaningful forms of voice become all the more important.

The profusion of voice online, too, has paradoxical consequences. "Like" buttons and the ability to "mute" other users allow a person to produce changes in their own perception of the community without engaging in collective decision or producing collective change. The algorithms that process such inputs do not show their workings in public view. Thus such buttons seem to serve as affective pressure valves in the guise of effective acts, providing users the ability to alter their own experience but not enact collective change.

The dominant patterns of online platforms and communities have lowered the costs of certain kinds of exit and certain kinds of voice beyond what Hirschman could have imagined for the organizations of the early 1970s. Yet we find that a type of voice he could take nearly for granted in the fraternal organizations and social clubs of his time is rare online: effective voice. While participants can attempt to persuade, shame, and annoy power-holders, platforms do not provide a process by which organized users can vote a moderator out of the role, for instance. Administrators with the power to censor, remove, and ban fellow users do not have to submit to transparency requirements or judicial oversight. When he famously argued that "code is law," Lawrence Lessig (2006) stressed the need for scrutiny over the kinds of regulation that community technologies impose online. The absence of effective voice has gone largely unnoticed.

## 3. Mechanisms for effective voice

What kinds of effective voice would best suit the contexts of online communities? Given the specific dynamics of agency and control described above, familiar effective mechanisms are not likely to function online in the same ways they do offline. In this section, we suggest avenues for introducing effective voice. We propose these as invitations for future research, as well as opportunities for platform designers seeking to deepen user loyalty through participation, and demands that users themselves might pose to platform companies. While many of these mechanisms resemble familiar practices in offline organizations and polities, we stress the distinctive forms they might take online.

In addition to Hirchman and his successors, our proposals draw guidance from the political scientist Elinor Ostrom and other scholars in her area of institutional analysis and design. This interdisciplinary research network has studied and analyzed the techniques by which communities around the world co-manage their shared resources, often for many generations. Research in the Ostrom community has emphasized participant voice and power, not just as a form of complaint but as a more multifaceted, and necessary, feature of institutional life (Kiser and Ostrom 2000; Ostrom, Walker, and Gardner 1992; DeCaro 2018). Given the limits of Hirschman's focus on dissatisfaction in the context of large, hierarchical institutions, Ostrom's attention to more participatory sorts of community contributes to thinking about exit and voice online, where institutional structures may be more fluid, the centers of authority less fixed, and the possible modes of participation more varied.

We first discuss mechanisms related to allocating authority and holding those with authority accountable, up and down the ladders of participation; these provide some assurance that power-holders can claim, in the idiom of political theory, the consent of the governed. Next we suggest forms of effective voice that are also collective, counteracting certain tendencies to individualize and isolate online users' complaints. The third set of mechanisms relates to guiding the inevitable evolution of communities, which might involve transitioning among various means of participation. Together these represent a palette of options for further research and experimentation, not a single proposal or formula ready-made for implementation.

**Mechanisms for authority and accountability**

What checks on administrator authority should users expect? Because consolidated control over online communities is so easy and ubiquitous, it would seem all the more necessary for participants to hold accountable those who hold that control, particularly when exit would be costly. Meanwhile, systems that permit users to grow into positions of leadership can stretch the ladder of participation to reach more community members. Therefore, we suggest effective mechanisms for defining and assigning authority in online communities.

*Make a community's rules binding on administrators.* The possibility of effective voice depends on the existence of rules that are binding on all participants, whatever their role. This might require rules that are enforced computationally, placing hard limits on behavior, but these rules might also be just as effective in the form of norms and agreements enforced by social expectations. On Wikipedia, for instance, the main mechanisms for keeping the various administrator roles accountable emanate more from user norms than platform design (Joyce, Pike, and Butler 2013). Conversely, blockchain protocols frequently bind administrators and founders with algorithmically mediated consensus mechanisms (De Filippi and Wright 2018).

*Provide participatory processes for the selection and removal of those in authority.* The means of conferring authority on leaders is a basic feature of modern participatory governance (Blondel 2014), whether in governments, corporations, or civic associations, but those means are largely unavailable to users in online communities. Enabling users to periodically select (and remove) their administrators would be a simple and familiar way of introducing effective voice. Algorithmic processes may also aid in introducing more dynamic forms of selection, such as by automatically assembling random juries of moderators or weighting users' voting power based on a calculus of reputation or participation in the community. Effective voice need not limit itself to delegating representatives, but it is a start.

**Mechanisms for collective action**

Self-organized collective action is an archetypal social dilemma (Olson 2003), and online manifestations of it—peer production, crowdsourcing, and online community—succeed or fail on their ability to attract small commitments from large populations of users. Mobilizing collective action may become easier, and ultimately more beneficial to platforms and users, if groups can organize to achieve binding goals through transparent processes.

*Provide binding mechanisms for aggregating user demands.* The most common online collective action mechanisms—upvotes, follows, flags, and reports—have uncertain effects on the social sphere. Moderators of Facebook Groups, for example, are not obliged to consider users' Reactions in determining what content is "pinned" to the top of a group's feed, nor to indicate how Reactions are aggregated into moderation decisions. Online petitions are similar: however persuasive or popular, they place no direct obligations on the petitioned organization. Platforms could, however, introduce a known threshold after which a petition or referendum would become binding, akin to the commitment on Madrid's CONSUL platform that a proposal supported by one percent of the city's population must be considered by the city council (Stempeck 2020).

*Let communities self-organize internal blocs.* As communities grow and mature, adapt and integrate, it is common for them to be host to small ecosystems of emergent internal structures such as unions, advocacy groups, or parties. Institutional scholars have argued that such "polycentric" structure improves a system's adaptability and resilience (McGinnis 2000; Ostrom and Janssen 2004)—as some online communities have discovered already. In amateur-run Minecraft game communities, the popular "Factions" plugin adds a competitive teaming dimension to the game by crossing administrative hierarchy with division into cultural tribes. Wikipedia editors can use "project" pages and areas of specialty on the site to organize around shared interests in fighting vandalism, supporting the bureaucracy, or improving content. The Taiwanese government's vTaiwan project utilizes Polis, software that organizes participants' into emergent interest groups using artificial intelligence (Stempeck 2020). Whether directed by users or algorithms, group formation can aid and amplify other forms of effective voice.

*Layer multiple mechanisms that allow for diverse forms of input.* On a petition platform like Change.org, users have two forms of voice: voting on a petition, which is efficient but not very expressive, and creating one, which is more demanding but allows for richer expression. The theory of institutional diversity proposes that the most resilient organizations overlay a wide variety of mechanisms (Ostrom 2006), and Hirschman (1970) repeatedly dashes hopes of any single solution to complex governance challenges. A further form of layering can be found in the "liquid democracy" techniques adopted by tech companies and Internet-based political parties, which modify direct democracy by allowing users to either vote directly on any proposal or delegate their votes on different subjects to topic experts they trust (Hardt and Lopes 2015). Diverse forms of effective voice can empower diverse participants, prevent a few from consolidating power through a unitary mechanism, and mitigate what Kelty (2019) calls "formatted" participation—in which what is straightforwardly computable can take priority over the wider range of human expression.

**Mechanisms for community change**

Online communities and their underlying platforms evolve over time. Research in the Ostrom community has seen consistently that the form of governance gains complexity with the complexity of the setting (Frey and Sumner 2019; Ostrom and Whitaker 1973). The growth of the early community-led news website Slashdot motivated several redesigns of the site's moderation system, from manual moderation by the founder to ever-expanding methods of delegating the task to users, ultimately giving reputable users considerable opportunities for effective voice over Slashdot's content (Ganley and Lampe 2009). And the developers of the Python programming language, after decades under the leadership of its "benevolent dictator," undertook a months-long, wide-ranging search for a replacement, eventually opting by referendum for an elected board structure; the process employed the community's tool for evaluating software changes, the Python Enhancement Proposals system (Edge 2018). More dedicated tooling could further facilitate effective voice not just within community rule-sets but over those rule-sets as well (Zhang, Hugh, and Bernstein 2020; Schneider et al. 2020), which in turn may increase the sustainability and adaptability of a community (Frey, Krafft, and Keegan 2019).

*Design for participation in structural change.* With appropriate planning and tooling, platforms can facilitate the evolution of their communities in the direction of effective voice. Prescribed trigger events offer one mechanism. Reddit considers a community "abandoned" if its moderators have been absent for a certain period of time, and others can then request the moderator role. Inversely, when a community reaches a certain number of members or level of activity, the platform might require a referendum on whether to keep the current governance model or adopt a new one. Enabling this kind of evolution would also require platforms to support diverse governance structures at the level of software, such as an "oligarchy mode" or a "consensus mode." Just as constitution-level changes often require a referendum in otherwise representative political systems, similarly structural changes seem like an especially appropriate use-case for online effective voice. If opportunities for democracy should be anywhere, they should be here.

## Conclusion

We have sought to contribute to ongoing debates about governance in (and of) online communities by amending Albert O. Hirschman's famous exit-voice distinction with the concept of effective voice. While forms of what we call affective voice are widespread online—and can be quite powerful as they circulate across networks—we argued that direct participation in governance through effective voice is largely unavailable, despite the advantages it could confer. We then presented a series of mechanisms for effective voice that designers and researchers might consider adopting, and that users might demand.

In doing so, we follow the spirit of Hirschman (1970)'s sympathy for forms of agency that foster loyalty and deeper participation over a simple exit. We suspect that introducing avenues for effective voice can enable more accountable and responsive online communities than those that prevail today. Yet we also follow Hirschman in recognizing that voice in general, or any particular variety of it, is not sufficient to furnish a successful organizational culture. Any given practice may not fit squarely within either concept. Mechanisms for voice—effective or otherwise—should exist as part of a wider range of options, including that of exit.

The binary clarity of online exit options approximate some features of effective voice, as community managers can see declining membership as an ongoing referendum on their behavior. But online exit is not always as easy as it may appear, and even when it is, effective voice can offer additional benefits. Further, for those concerned about the effects of online communities on civic life, effective voice may have benefits in fostering more vibrant cultures of democratic participation, online and off (Campante, Durante, and Sobbrio 2018; C. Kelty et al. 2015; T. Graham and Witschge 2003).

We do not, however, contend that a profusion of mechanisms for effective voice would represent an unmitigated good. Greater reliance on mechanistic and computational decision-making could tempt communities to neglect the work of consensus-building and the cultivation of a shared culture. Transparent governance mechanisms could also invite manipulation from users seeking to undermine the process altogether. Again, we hope that best practices and good habits would emerge from experimentation.

Our case for effective voice may appear idealistic or impractical. Do users really want the responsibility of governing the various spaces of their online lives? Facebook, for instance, instituted a voting process in 2009 surrounding a proposed change to the platform's policies; only about 600,000 users cast votes, short of the 60 million—30 percent of active users at the time—that would have made the votes binding as effective voice (Gaudin 2009). Was the cause of poor turnout a matter of laziness, a lack of awareness, an absence of compelling stakes, an expression of the arcane nature of terms of service, or a dislike for both choices offered? As in political and civic organizations, the answer to the question of whether people would participate in more direct levers of online governance, given the chance, is surely "it depends"—on the context, on the levers, on the perception of whether the difference will make a difference.

Another variable lies outside the purview of the online communities themselves—the ownership of the firms that host those communities' platform substrates. Community-level effective voice is surely possible regardless of the corporate regime that controls the software, but perhaps it stands to reason that democratically governed peer-production projects like Debian and Wikipedia operate under the umbrella of nonprofit organizations, and that Loomio, a community decision-making platform, is the work of a worker-owned cooperative. Effective voice need not wait for the widespread arrival of platform cooperativism (Scholz and Schneider 2016), but democratic platforms might be especially likely to recognize their users as potential democratic agents.

We hope that the avenues of experimentation we suggest here will entice designers, researchers, and users to further explore the possibilities of effective voice online, which could evolve more rapidly and creatively than the forms of participatory governance that offline social practices have enabled. We believe the trouble of experimentation will prove worthwhile. A decade after the publication of *Exit, Voice, and Loyalty*, Hirschman observed that, in the struggles of social life, often "the cost of voice turns into a benefit" (Hirschman 1980, 431).

## Acknowledgements

The authors claim equal effort. They are grateful for feedback on early versions from participants in the Community Data Science Lab and the Metagovernance Seminar. This work was supported by NSF RCN 1917908: "Coordinating and Advancing Analytical Approaches for Policy Design" and NSF GCR 2020751: "Jumpstarting Successful Open-Source Software Projects With Evidence-Based Rules and Structures," as well as an Open Society Fellowship from the Open Society Foundations.

EFFECTIVE VOICE 32

EFFECTIVE VOICE 35